\documentstyle[aps,prl]{revtex}

\draft

\begin{document}

\onecolumn

\title{Optomechanical tailoring of quantum fluctuations}

\author{S. Mancini}
\address{INFM and Dipartimento di Fisica, Universit\`a di Milano,
Via Celoria 16,
20133, Italy}

\author{H. M. Wiseman }
\address{School of Science, Griffith University, Nathan, Brisbane,
Queensland 4111 Australia}

\date{\today}

\maketitle

\begin{abstract}
We propose the use of feedback mechanism to control the level of quantum
noise
in a radiation field emerging from a pendular Fabry-Perot cavity.
It is based on the possibility to perform quantum-nondemolition
measurements by means of optomechanical
coupling.
\end{abstract}


\section{Introduction}

Optomechanical coupling between light and (movable) mirrors gives
rise to a variety of intriguing effects.
A wide class of quantum states coming from such coupling has been
investigated in Ref. \cite{bose97}.
Furthermore, due to the recent technological developments in optomechanics,
this area is now becoming experimentally accessible \cite{exp} and quite
interesting.
Usually in such a systems,
a mirror moves in response to radiation pressure exerted by the light.
This leads to an intensity-dependent phase shift for the light field
equivalent to an optical Kerr
effect \cite{kerr}.
Hence, among the others, this system shows bistable behaviour
\cite{bistab}, and quantum noise
reduction effects \cite{man94,fab94};  moreover, it can be used to realize
quantum-nondemolition
(QND) measurement on the light field \cite{jac94}.
Some of these apects have been recently studied also in the spatial domain
\cite{spatial}.

On the other hand, the use of feedback to control the noise in optical
system has been long used. Classical theory is often adequate to
describe such feedback.  However,
the advent of nonclassical light (in particular squeezed light),
required a more
accurate treatment. A number of useful
approaches exist, using linearized quantum
Langevin equations \cite{YamImoMac86,Sha87} and, more recently,
exact quantum trajectories as well \cite{fbtheo}.
In particular, it has been shown that it is
possible to use QND-mediated
feedback to get squeezing, both extracavity \cite{YamImoMac86,Sha87}
and intracavity \cite{sqfb}.

Along this line, here we investigate the possibility of controlling the
light statistics
via optomechanical coupling and a consequent feedback action.

The paper is organized as follows:
in Section II we present the measurement model, in Section III we study the
light field
dynamics, and the spectra are shown in Section IV. Then, in Section V we
introduce the feedback
action and show its influence on the light statistics.
Section VI concludes with a numerical result.

\section{The Measurement Model}

The system we wish to study consists of a Fabry-Perot cavity with
a partially transmitting mirror on one side and a perfectly reflecting, and
oscillating, mirror on
the other side. We imagine the latter coated on the surface of a
piezoelectric crystal and then we shall adopt the model of Ref.
\cite{pin95}. In such a case the variation of the crystal length
can be measured by an
electric circuit. Hence, in reality, the system is composed of three
different coupled subsystems,
the light, the crystal and the electric circuit.
The light field is coupled to the crystal length by the radiation pressure
force, while the
crystal and the electric circuit are coupled by piezoelectric effect.

Concerning the light field, the system is equivalent to the usual
Fabry-Perot cavity except that
the cavity detuning may vary because of the crystal elasticity. The
dynamics of the intracavity
field $a$ in a rotating frame is  described by
\begin{equation}\label{adyn}
\frac{da(t)}{dt}=-\left[\gamma-i\tau^{-1}\Psi(t)\right]a(t)+\sqrt{2\gamma}
\,a_{\rm in}(t)\,,
\end{equation}
where $\tau$ is the cavity round trip time and $\gamma$ is the
amplitude
damping rate
of the cavity (equal to the transmissivity of the fixed mirror divided
by $\tau$).
Furthermore, $a_{\rm in}$ is the incoming field, and it is related to the
intracavity and
outgoing ones by the input-output relation
\begin{equation}\label{ainout}
a_{\rm out}(t)=\sqrt{2\gamma} a(t) - a_{\rm in}(t)\,.
\end{equation}
The cavity detuning $\Psi/\tau$ depends linearly on the variation $x(t)$ of the
crystal length
\begin{equation}\label{Psi}
\Psi(t)=\tau\Delta +2k_0x(t)\,,
\end{equation}
and couples the mechanical motion of the crystal to the field.
Here, $k_0=\omega_0/c$ is the incoming field wave number, while
$\Delta$ is the cavity detuning in the absence of the mirror's motion.

The mechanical response of the crystal is described by the variation $x(t)$
of the crystal length.
For small displacements, the linear response theory \cite{landau} can be
invoked
\begin{equation}\label{xF}
x(\omega)=\chi_0(\omega)\left[F_R(\omega)+\zeta
Q(\omega)+F_T(\omega)\right]\,,
\end{equation}
where $\chi_0(\omega)$ is the mechanical susceptibility
\begin{equation}\label{chi0}
\chi_0(\omega)=\frac{1/m}{\omega_m^2-\omega^2-i\omega\gamma_m}\,,
\end{equation}
of the (movable mirror) crystal which has mass $m$, oscillates at frequency
$\omega_m$,
and it is damped at rate $\gamma_m$.

The first force acting on the mirror, $F_R$,
is the radiation pressure force given by
\begin{equation}\label{FR}
F_R=2\hbar k_0 a^{\dag}a/\tau\,.
\end{equation}
This has the simple interpretation of the momentum change of a photon under
reflection, times the number of photons, divided by the round-trip
time.

The electric circuit induces the presence of an electric charge $Q$ on one
side of the crystal
and $-Q$ on the other side (the current $I$ being the time derivative of $Q$,
i.e. $I(\omega)=-i\omega Q(\omega)$).
These charges generate a variation of the crystal length proportional to
$Q$. This effect is
equivalent to a force applied on both sides of the crystal, with opposite
sign and absolute value
$\zeta Q$, where $\zeta$ is a constant depending on the crystal
characteristics \cite{piezo}.

Finally, $F_T$ corresponds to the Langevin force describing the coupling of
the mechanical motion
with a thermal bath. Its spectrum \cite{defS} is given by
\begin{equation}\label{SFT}
{\cal S}_{F_T}=
2m \gamma_m k_B T\,,
\end{equation}
where $T$ is the temperature of the thermal bath and $k_B$  the Boltzmann
constant.
This relation is valid when $T$ is such that $k_B T\gg \hbar \omega_m$.
Otherwise, one should bring some corrections to the dynamics of the
mechanical oscillator
\cite{diosi}.

Now, the piezoelectric crystal is equivalent to a capacitance $C$ and a
voltage generator
in series. A length variation induces a polarization inside the crystal
which generates a potential
difference between the two sides of the crystal. This voltage is
proportional to the length
variation ($V=-\zeta x$) and thus, a measurement of this voltage allows one
to monitor the crystal
motion, and to measure the amplitude quadrature of the intracavity field
(or equivalently the
radiation pressure force).

The measured signal is the current $I^{\rm out}$ going out of the system by a
coaxial line;
a counterpropagating current $I^{in}$ enters the system by the same line.
This latter corresponds
to the Nyquist noise of the characteristic resistance $R$ of the line.
The input-output relation for the current is \cite{duffin}
\begin{equation}\label{Iinout}
I=I^{\rm in}+I^{\rm out}\,,
\end{equation}
and the noise spectrum of the current $I^{\rm in}$
entering the system is given by
\begin{equation}\label{SIin}
{\cal S}_{I^{\rm in}}=\frac{k_BT}{2R}\,.
\end{equation}
The voltage at the entrance of the circuit is related to the currents by
two equations,
one for the line and one for the circuit
\begin{eqnarray}
V(\omega)&=&R\left[I^{\rm in}(\omega)-I^{\rm out}(\omega)\right]\,,
\label{V1}\\
V(\omega)&=&Z_0(\omega)I(\omega)-\zeta x(\omega)\,,
\label{V2}
\end{eqnarray}
where $Z_0$ is the (purely imaginary) impedance of the circuit including the
capacitance $C$ of the crystal, and $-\zeta x$ is the voltage generated by
the
piezoelectric effect.

From Eqs.(\ref{Iinout}), (\ref{V1}), (\ref{V2}),
and with the aid of Eq.(\ref{xF}) it is possible to get
the output, measured current as
\begin{equation}\label{Iout}
I^{\rm out}(\omega)=
\frac{R-Z(\omega)}{R+Z(\omega)}I^{\rm in}(\omega)
+\frac{\zeta\chi_0(\omega)}{R+Z(\omega)}
\left[F_T(\omega)+F_R(\omega)\right]\,,
\end{equation}
where $Z(\omega)$ is an effective impedance
\begin{equation}\label{Z}
Z(\omega)=Z_0(\omega)-i\frac{\zeta^2}{\omega}\chi_0(\omega)\,.
\end{equation}
We  see that the measured current provides information about the radiation
pressure force.

\section{The Light Field Dynamics}

Following Ref. \cite{pin95},
to study the light field dynamics we exploit the linearisation procedure.
The equation for the mean field, which follows from Eq.~(\ref{adyn}),
is
\begin{equation}\label{sseq}
0=-\left[\gamma-i\psi/\tau\right]\alpha+\sqrt{2\gamma}\,\alpha_{\rm in}\,,
\end{equation}
where $\psi$ represents the steady state detuning.
The solution reads
\begin{equation}\label{alpha}
\alpha=\frac{\sqrt{2\gamma}}{\gamma-i\psi/\tau}\alpha_{\rm in}
\,.
\end{equation}
For the sake of simplicity, we shall assume, from now on, $\alpha_{\rm in}$ to
be real. After linearisation, we get the equation for the quantum
fluctuations
\begin{equation}\label{daeq}
\frac{d}{dt}\,\delta a(t)=-\left[\gamma-i\tau^{-1}\psi\right]\,\delta a(t)
+i\alpha\,\delta\Psi(t)/\tau+\sqrt{2\gamma}\,\delta a_{\rm in}(t)\,.
\end{equation}

Now, by defining the generic quadrature
\begin{equation}\label{quadef}
X_{\theta}=\left(ae^{-i\theta}+a^{\dag}e^{i\theta}\right)\,,
\end{equation}
the radiation pressure force (\ref{FR}) becomes
\begin{equation}\label{FRlin}
F_R=2\hbar k_0\tau^{-1}\left[|\alpha|^2+\alpha\delta
a^{\dag}+\alpha^*\delta a\right]
=2\hbar k_0\tau^{-1}|\alpha|\left(|\alpha|+\delta X_{\arg\alpha}\right)\,,
\end{equation}
so, it is proportional to the fluctuations of a quadrature
determined by the stationary value of the intracavity field.

The equation of motion (\ref{daeq}), can be rewritten as
\begin{equation}\label{dXeq}
\frac{d}{dt}\,\delta X_{\theta}(t)=-\gamma\,\delta X_{\theta}(t)
-\tau^{-1}\psi\,\delta X_{\theta+\pi/2}(t)
+2|\alpha|\tau^{-1}\sin\left(\theta-\arg\alpha\right)\,\delta\Psi(t)
+\sqrt{2\gamma}\,
\delta X^{\rm in}_{\theta}(t)\,,
\end{equation}
where the spectrum of the input fluctuations is
${\cal S}_{X^{\rm in}_{\theta}}=1$
for coherent light.
We immediately see that the measured
quadrature $\theta=\arg\alpha$ is not affected by
the measurement process (i.e. being a QND observable)
provided to have $\psi=0$ \cite{jac94}.
So, from now on we assume such a condition, which also
implies $\alpha$ to be real (i.e. $\arg\alpha=0$).
Thus, in the frequency domain, the solution of Eq.(\ref{dXeq}) reads
\begin{equation}\label{dXth}
\delta X_{\theta}(\omega)=
\frac{2\alpha\sin\theta}{(\gamma-i\omega)\tau}\,
\delta\Psi(\omega)
+\frac{\sqrt{2\gamma}}{(\gamma-i\omega)}\,\delta
X^{\rm in}_{\theta}(\omega)\,.
\end{equation}

Any real measurement device must give a read-out and in this case it
is the current $I^{\rm out}$. Hence the fluctuations of this current
is directly related to the
fluctuations
of the quadrature $X_0$. Specifically, by using Eqs. (\ref{Iout}), and
(\ref{FRlin}), we have
\begin{equation}\label{dIout}
\delta I^{\rm out}(\omega)=
\alpha_{\rm in}{\cal G}(\omega) \delta X^{\rm
in}_{0}(\omega)+I_T(\omega)\,,
\end{equation}
where the gain ${\cal G}$, and the thermal current $I_T$, are given by
\begin{equation}\label{G}
{\cal G}(\omega)=2\hbar k_0\frac{1}{(\gamma-i\omega)\tau}
\frac{\zeta\chi_0(\omega)}{R+Z(\omega)}\,,
\end{equation}

\begin{equation}\label{IT}
I_T(\omega)=
\frac{R-Z(\omega)}{R+Z(\omega)}I^{\rm in}(\omega)
+\frac{\zeta\chi_0(\omega)}{R+Z(\omega)}
F_T(\omega)\,.
\end{equation}
In the case of equal mechanical and electric temperatures, the spectrum of
the thermal current can
be easily calculated by means of Eqs.(\ref{SIin}), (\ref{SFT}), and
(\ref{Z}), and the result
\begin{equation}\label{SIT}
{\cal S}_{I_T}=\frac{k_B T}{2R}\,,
\end{equation}
as required for thermal equilibrium.

\section{The Light Spectrum}

In order to derive the light spectra we first calculate the
spectrum ${\cal S}_{\Psi}$ and the correlations
${\cal C}_{\Psi\,X^{\rm in}_{\theta}}$, ${\cal C}_{\Psi\,I_T}$ \cite{defC}.
To this end, we remind
that the current $I$ is the time derivative of the charge $Q$
(i.e. $I(\omega)=-i\omega Q(\omega)$)
appearing in Eq.(\ref{xF}), and then we eliminate the electric variable
in Eqs.(\ref{xF}), and (\ref{V2}), obtaining
\begin{equation}\label{xFnew}
x(\omega)=\chi(\omega)\left[
F_R(\omega)+F_T(\omega)+i\frac{\zeta}{\omega}\frac{2R}{R+Z_0(\omega)}I^{\rm in}
(\omega)\right]
\end{equation}
where
\begin{equation}\label{chi}
\frac{1}{\chi(\omega)}=
\frac{1}{\chi_0(\omega)}-i\frac{\zeta^2}{\omega}\frac{1}{R+Z_0(\omega)}\,.
\end{equation}
This means that the mechanical response to the radiation pressure is
modified by the coupling.
As a consequence it will be
\begin{equation}\label{dPsi}
\delta \Psi(\omega)=
2k_0\chi(\omega)\left[
2\hbar k_0 \tau^{-1}|\alpha| \delta X_{0}(\omega)
+F_T(\omega) +i\frac{\zeta}{\omega}\frac{2R}{R+Z_0(\omega)}I^{\rm in}(\omega)
\right]\,.
\end{equation}
Now by considering Eqs.(\ref{dPsi}), (\ref{IT}), (\ref{dXth}),
(\ref{SIin}), and (\ref{SFT})
we can calculate the desired quantities
\begin{equation}\label{SPsi}
{\cal S}_{\Psi}(\omega)=
4k_0^2\mid\chi(\omega)\mid^2
\left[
4\hbar^2 k_0^2|\alpha|^2\frac{2\gamma}{(\gamma^2+\omega^2)\tau^{2}}
+2m \gamma_m k_BT
+\frac{\zeta^2}{\omega^2}\frac{2R}{\mid R+Z_0(\omega)\mid^2}k_BT
\right]\,,
\end{equation}

\begin{equation}\label{CPsiX}
{\cal C}_{\Psi\,X^{\rm in}_{\theta}}=
4\hbar k_0^2\alpha\chi(\omega)\frac{\sqrt{2\gamma}}{(\gamma-i\omega)\tau}
e^{i\theta}\,,
\end{equation}
and
\begin{equation}\label{CPsiIT}
{\cal C}_{\Psi\,I_T}=
2k_0\chi(\omega)\left[
\frac{\zeta\chi_0^*(\omega)}{R+Z^*(\omega)}2m\gamma_m
+i\frac{\zeta}{\omega}\frac{1}{R+Z_0(\omega)}\frac{R-Z^*(\omega)}
{R+Z^*(\omega)}
\right]k_BT\,.
\end{equation}

Now, from Eq.~(\ref{dXth}) the intracavity spectrum can be written as
\begin{eqnarray}\label{Sintra}
{\cal S}_{X_{\theta}}&=&
\frac{4\alpha^2\sin^2\theta}{
(\gamma^2+\omega^2)\tau^2}
{\cal S}_{\Psi}(\omega)
+\frac{2\gamma}{\gamma^2+\omega^2}
\nonumber\\
&+&\frac{2\sqrt{2\gamma}\alpha\sin\theta}
{(\gamma^2+\omega^2)\tau}
2{\rm Re}\left\{{\cal C}_{\Psi\,X^{\rm in}_{\theta}}\right\}\,.
\end{eqnarray}
Finally, the output spectrum can be derived by using the input-output relation
(\ref{ainout}).
The result is
\begin{eqnarray}\label{Sout}
{\cal S}_{X^{\rm out}_{\theta}}&=&
\frac{8\gamma \alpha^2\sin^2\theta}
{(\gamma^2+\omega^2)\tau^2}{\cal S}_{\Psi}(\omega)
+1
\nonumber\\
&+&\frac{2\sqrt{2\gamma}\alpha\sin\theta}
{(\gamma^2+\omega^2)\tau}
2{\rm Re}\left\{(\gamma+i\omega){\cal
C}_{\Psi\,X^{\rm in}_{\theta}}\right\}\,.
\end{eqnarray}
It is immediately apparent that the output spectrum (\ref{Sout})
remains equal to one for $\theta=0$ [a fact obvious from
Eq.~(\ref{dXth}) and used in
Eq.~(\ref{SPsi})], while it takes
its maximum value for $\theta=\pi/2$.
This is due to the back action of the mirror noise, which in the
regime we consider
is dominated by thermal noise.
Furthermore, calculating Eq.(\ref{Sout}) we neglected 
the thermal photons generated by the mirror motion. 
This is reasonable at optical frequencies ($\omega_0$)
since the mechanical frequency ($\omega_m$) 
is much smaller \cite{man94}.

Notice, that in addition to Ref. \cite{pin95} we explicitely
calculated the spectrum for the detuning and for a generic quadrature.

\section{The Feedback Loop}

We now assume  that the cavity mode
is controlled by modulating the input field linearly with the
measured current.
This can be achieved by using electro-optic modulators \cite{tau95}.
To model this feedback mechanism, we use the quantum Langevin approach
rather than the quantum trajectory approach \cite{fblan}.
Moreover, to simplify the problem, we assume that the feedback is Markovian,
that is the response function of the feedback loop is approximately flat
from zero to a frequency
much larger than that of the mirror or of the electric circuit.
This is not unreasonable, as electronic and
electro-optic devices can respond on time
scales much shorter than $10^{-6}$s.
Hence, the feedback can be modeled by changing the input beam
according to the following rule
\begin{equation}\label{aintil}
a_{\rm in}(t)=\alpha_{\rm in}+\delta a_{\rm in}(t)
\quad\to\quad
{\widetilde a}_{\rm in}(t)=\alpha_{\rm in}+\delta a_{\rm in}(t)-\lambda
\delta I^{\rm out}(t)\,.
\end{equation}
In terms of the quadrature fluctuation operator
\begin{equation}\label{dXintil}
\delta X^{\rm in}_{\theta}(t)
\quad\to\quad
\delta {\widetilde X}^{\rm in}_{\theta}(t)=\delta X^{\rm in}_{\theta}(t)
-\left(\lambda e^{-i\theta}+\lambda^* e^{i\theta}\right) \delta I^{\rm
out}(t)\,.
\end{equation}
Here, the value of $\lambda$ is related to the practical way
of realize the feedback action, i.e., to the electro-optics modulators
capabilities.

With the choice of $\lambda$ real,
the quadrature Eq.(\ref{dXth}) is modified as follows
\begin{eqnarray}\label{dXtil}
\delta {\widetilde X}_{\theta}(\omega)&=&
\frac{2\alpha\sin\theta}{(\gamma-i\omega)\tau}\delta\Psi(\omega)
+\frac{\sqrt{2\gamma}}{(\gamma-i\omega)}\delta X^{\rm in}_{\theta}(\omega)
\nonumber\\
&-&2\lambda\frac{\sqrt{2\gamma}}{\gamma-i\omega}
\left[\cos\theta
+f(\omega)\sin\theta\right]
\left[\alpha_{\rm in}{\cal G}(\omega) \delta X^{\rm in}_{\varphi}(\omega)
+I_T(\omega)\right]\,,
\end{eqnarray}
where
\begin{equation}\label{fom}
f(\omega)=8\hbar k_0^2 \frac{\alpha^2\chi(\omega)}
{(\gamma-i\omega)\tau^2}\,,
\end{equation}
is the feedback term coming from the modified detuning fluctuations,
i.e. the term proportional to $\lambda$ when the replacement (\ref{dXintil})
is made in Eq.(\ref{dPsi}).

The intracavity spectrum then becomes
\begin{eqnarray}\label{SXthfb}
{\cal S}_{{\widetilde X}_{\theta}}={\cal S}_{X_{\theta}}
&-&4\lambda\frac{2\gamma}{\gamma^2+\omega^2}
{\rm Re}\left\{
\alpha_{\rm in}{\cal G}^*(\omega)e^{-i\theta}
\left[\cos\theta+f^*(\omega)
\sin\theta\right]\right\}
\nonumber\\
&-&4\lambda\frac{\sqrt{2\gamma}\alpha}{(\gamma^2+\omega^2)\tau}
{\rm Re}\left\{
\left[\sin\left(2\theta\right)+2f^*(\omega)
\sin^2\theta\right]
\left[
\alpha_{\rm in}{\cal G}^*(\omega){\cal C}_{\Psi\,X^{\rm in}_{\varphi}}(\omega)
+{\cal C}_{\Psi\,I_T}(\omega)\right]
\right\}\nonumber\\
&+&4\lambda^2\frac{2\gamma}{\gamma^2+\omega^2}
\left[\alpha_{\rm in}^2\mid{\cal G}(\omega)\mid^2
+\frac{k_BT}{2R}\right]
\left|\cos\theta+f(\omega)\sin\theta\right|^2
\,.
\end{eqnarray}

In view of Eqs.(\ref{dXintil}), (\ref{dXtil}),
the input-output relation (\ref{ainout}) gives
\begin{equation}\label{dXouttil}
\delta {\widetilde X}_{\theta}^{\rm out}=\delta X_{\theta}^{\rm out}
-2\lambda\left[
\frac{\gamma+i\omega }{\gamma-i\omega }\cos\theta
+\frac{2\gamma f(\omega)}{\gamma-i\omega}\sin\theta
\right]
\left[\alpha_{\rm in}{\cal G}(\omega)\delta X_{0}^{\rm in}(\omega)
+I_T(\omega)\right]\,.
\end{equation}
Finally, the output spectrum in the presence of feedback action is
\begin{eqnarray}\label{SXoutfb}
{\cal S}_{{\widetilde X}_{\theta}^{\rm out}}=
{\cal S}_{X_{\theta}^{\rm out}}
&-&4\lambda{\rm Re}\left\{
\alpha_{\rm in}{\cal G}^*(\omega)e^{-i\theta}
\left[\cos\theta
+\frac{2\gamma f^*(\omega)}{\gamma-i\omega}\sin\theta
\right]\right\}
\nonumber\\
&-&4\lambda{\rm Re}\left\{
\frac{\sqrt{2\gamma}\alpha}{\tau}
\left[
\frac{\sin\left(2\theta\right)}{\gamma+i\omega}
+\frac{4\gamma f^*(\omega)\sin^2\theta}{\gamma^2+\omega^2}
\right]
\left[\alpha_{\rm in}{\cal G}^*(\omega) \, {\cal C}_{\Psi\,
X^{\rm in}_{0}}(\omega)
+{\cal C}_{\Psi\, I_T}(\omega)\right]
\right\}
\nonumber\\
&+&4\lambda^2\left[
\alpha_{\rm in}^2\mid {\cal G}(\omega)\mid^2+\frac{k_BT}{2R}\right]
\left|
\frac{\gamma+i\omega}{\gamma-i\omega}\cos\theta
+\frac{2\gamma f(\omega)}{\gamma-i\omega}\sin\theta
\right|^2
\,.
\end{eqnarray}
For the quadrature $X_{0}$ the above expression reduces to
\begin{equation}\label{SXphifb}
{\cal S}_{{\widetilde X}^{\rm out}_{0}}=
\Big| 1-2\alpha_{\rm in}\lambda{\cal G}(\omega)
\Big|^2+4\lambda^2\frac{k_BT}{2R}\,.
\end{equation}

For any given frequency $\omega$, it is possible to choose some value of
feedback strength
$\lambda$ such that this expression will be less than one, providing
the real part of ${\cal G}(\omega)$ is positive.
That is, the feedback enables one to produce a sub-shot-noise output.
The optimum value for the feedback parameter results
\begin{equation}\label{lambopt}
\lambda_{\rm opt}=\frac{\alpha_{\rm in} {\rm Re}\left[{\cal G}(\omega)\right]}
{2\alpha_{\rm in}^2 \left|{\cal G}(\omega)\right|^2+\frac{k_BT}{R}}\,.
\end{equation}
For the amount of squeezing to be significant it is necessary for the
measurement gain to be mostly positive and also
large compared to the thermal noise:
\begin{equation}\label{cond2}
\alpha_{\rm in}\left|{\cal G}(\omega)\right|
\gg \sqrt{\frac{k_B T}{2R}} \,.
\end{equation}
The same conditions hold for the intracavity noise as can be seen from
Eq.(\ref{SXthfb}).

\section{Conclusion}

The main result, obtained in the preceding section, is that
using an electro-optic
feedback mechanism it is possible to reduce
the quantum fluctuations of a field quadrature below the shot-noise
limit. The amount of noise
reduction depends on the value
of the feedback parameter $\lambda$ as well as on the frequency $\omega$
determining the gain in the measurement process.
The best squeezing is obtained in the amplitude quadrature (with respect to
the input field) for which a perfect QND measurement is achievable.
On the other hand a
rotation of the ellipse of fluctuations \cite{gala}
could be devised at the output.
It is also worth noting that in the model we have elaborated, the added
thermal noise
(Eq.(\ref{SXphifb})) can be
controlled by means of the electric components (e.g. the resistance),
contrary to the case of
Refs. \cite{man94,fab94}, where thermal fluctuations lead to anavoidable
detrimental effects.

To better show the potentialities of our model, we consider a realistic
experimental situation.
First of all, to enhance the measurement efficiency,
we imagine we have a resonant electric circuit.
With self-induction of impedance $L$ the electric resonance frequency
$\omega_e$
is given by
\begin{equation}\label{omegae}
LC\omega_e^2=1\,,
\end{equation}
and we choose this to be almost equal to $\omega_m$.
The noncoupled electric impedance $Z_0$
is
\begin{equation}\label{Z0}
Z_0(\omega)=i\left(\frac{1}{C\omega}-L\omega\right)\,.
\end{equation}
Then, it is possible to define the electric damping and the piezoelectric
coupling frequency
\begin{equation}\label{electric}
\gamma_e=R/L\,,\quad
\Omega_e=\zeta\sqrt{C/m}\,.
\end{equation}
Now, we take for granted the values of parameters in Ref. \cite{pin95}.
The mirror parameters are:
mass $m=10^{-3}$ ${\rm g}$
(this value should be intended as an
effective value coming from the acoustic modes of the mirror
\cite{pin99}),
resonance frequency $\omega_m=10^{6}$ ${\rm s}^{-1}$,
quality factor $Q_m=\omega_m/\gamma_m=10^6$,
and piezoelectric coupling frequency $\Omega_e\approx 10^3$ ${\rm s}^{-1}$.
The cavity parameters are: bandwidth $\gamma=10^6$ ${\rm s}^{-1}$,
round trip time $\tau = 10^{-11}$s,
field wavelength  $0.5$ $\mu$m, and
incident power $P^{\rm in}=\hbar \omega_0 \alpha_{\rm in}^2=100$ mW.
The transmissivity of the fixed mirror should be the smallest possible since
the gain factor (\ref{G}) is inversely proportional to it.
Finally the electric quality factor $Q_e=\omega_e/\gamma_e\approx 10^6$.

In Fig.1 we display the spectrum of Eq.(\ref{SXphifb}) for different values
of temperature.
We see that the noise reduction occours in proximity of the mechanical
resonace frequency,
where the response of the meter is maximum.
It is worth noting that the squeezing bandwidth is approximately
proportional to the meter bandwidth $\gamma_m$, and whenever it is increased
the squeezing phenomenon becomes more sensible to the thermal noise since the
mechanical quality factor $Q_m$ decreases.

Finally, we would briefly discuss about
the quantum mechanical consistency of our
treatment. 
For the QND variable $X_{0}$, and its conjugate $X_{\pi/2}$
the commutation relations are preserved from the input to the output.
The same is not true in general for $X_{\theta}$ and $X_{\theta+\pi/2}$ 
with $\theta\ne 0$,
due to the fact that the mirror introduces 
beside a frequency dependent phase shift, an additional 
frequency-dependent damping \cite{man94,spatial}.
Hence, the canonical commutation relation 
$[X^{\rm out}_{0}(t),X^{\rm out}_{\pi/2}(t')] = 2 i \delta(t-t')$ holds, 
and it leads to the uncertainty relation 
\begin{equation}
{\cal S}_{X^{\rm out}_{0}}(\omega) {\cal S}_{X^{\rm out}_{\pi/2}}(\omega)
\ge 1\,,
\end{equation}
for all $\omega$. In reality, since there is a second output from our
system, the current $I^{\rm out}$, which measures $X^{\rm out}_{0}$,
we can demand a stronger inequality,
\begin{equation}\label{heis}
{\cal S}_{X^{\rm out}_{0} \mid I^{\rm out}}(\omega)
{\cal S}_{X^{\rm out}_{\pi/2}}(\omega) \ge 1\,.
\end{equation}
Here the first quantity is the spectrum of fluctuations in the
$\theta=0$ quadrature conditioned upon the measured current $I^{\rm out}$. 
For Gaussian statistics, which hold in the linearized regime we
are considering, this conditioned spectrum can be evaluated as
\begin{equation}
{\cal S}_{X^{\rm out}_{0} \mid I^{\rm out}}(\omega)
={\cal S}_{X^{\rm out}_{0}}(\omega)
 - \left|{\cal C}_{X^{\rm out}_{0}\,I^{\rm out}}(\omega)\right|^2
 \Big/{\cal S}_{I^{\rm out}}(\omega)\,.
\end{equation}
We have checked the relation (\ref{heis})
numerically for the regimes
of interest, both with and without feedback, and find that it is
always satisfied.

In conclusion, we have presented a realistic model, based on optomechanical
coupling,
for an active control
of the quantum noise of a radiation field.
The theory has been developed by using a semiclassical approach, which
nevertheless
guarantes quantum mechanical consistency for realistic situation.
Of course in a real situation minor detrimental effects could
appear;
we mention for example the non perfect reflectivity
of the rear mirror (which decreases the gain factor), or
the absorption and diffraction losses in the input coupler
(which introduce excess noise).
However, we think that the up-to-date technology is
mature enough
to envisage experimentally an optomechanical tailoring of quantum
fluctuations; this
could be useful in cavity based experiments,
or to produce nonclassical light; among other possible interesting
applications
we cite the reduction of the radiation pressure
noise in  gravitational interferometers \cite{grav}.

\begin{figure}
\caption{The quantity ${\cal S}_{{\tilde X}_{0}^{\rm out}}$ is plotted
vs $\omega$ for different values of temperature.
>From top to bottom, the curves correspond to $T=300$ K,
$T=70$ K, and $T=4$ K.
The values of the other parameters are set in the text, and
$\lambda$ is choosen according to Eq.(\ref{lambopt}).}
\end{figure}


\begin{references}

\bibitem{bose97}
S. Bose, K. Jacobs and P. Knight,
Phys. Rev. A {\bf 56}, 4175 (1997);
S. Mancini, V. I. Man'ko and P. Tombesi,
Phys. Rev. A {\bf 55}, 3042 (1997).

\bibitem{exp}
See e.g., I. Tittonen, G. Breitenbach, T. Kalkbrenner,
T. M\"uller, R. Conradt, S. Schiller, E. Steinsland,
N. Blanc and N. F. de Rooij,
Phys. Rev. A {\bf 59}, 1038 (1999);
Y. Hadjar, P. F. Cohadon, C. G. Aminoff, M. Pinard and
A. Heidmann, quant-ph/9901056;
P. F. Cohadon, A. Heidmann and M. Pinard,
Phys. Rev. Lett. {\bf 83}, 3174 (1999).

\bibitem{kerr}
P. Meystre, E. M. Wright, J. D. McCullen and E. Vignes,
J. Opt. Soc. Am. B {\bf 2}, 1830 (1985).

\bibitem{bistab}
A. Dorsel, J. C. McCullen, P. Meystre, E. Vignes
and H. Walther, Phys. Rev. Lett. {\bf 51}, 1550 (1983).

\bibitem{man94}
S. Mancini and P. Tombesi, Phys. Rev. A {\bf 49}, 4055 (1994).

\bibitem{fab94}
C. Fabre, M. Pinard, S. Bourzeix, A. Heidmann, E. Giacobino and S. Reynaud,
Phys. Rev. A {\bf 49}, 1337 (1994).

\bibitem{jac94}
K. Jacobs, P. Tombesi, M. J. Collett and D. F. Walls,
Phys. Rev. A {\bf 49}, 1961 (1994).

\bibitem{spatial}
S. Mancini, A. Gatti and L. A. Lugiato, preprint.

\bibitem{YamImoMac86}
Y. Yamamoto, N. Imoto and S. Machida,
Phys. Rev. A {\bf 33}, 3243 (1986).

\bibitem{Sha87}
J.M. Shapiro {\em et al},
J. Opt. Soc. Am. B {\bf 4}, 1604 (1987).

\bibitem{fbtheo}
H. M. Wiseman and G. J. Milburn,
Phys. Rev. Lett. {\bf 70}, 548 (1993);
H. Wiseman, Phys. Rev. A {\bf 49}, 2133 (1994).

\bibitem{sqfb}
H. M. Wiseman and G. J. Milburn, Phys. Rev. A
{\bf 49}, 1350 (1994).

\bibitem{pin95}
M. Pinard, C. Fabre and A. Heidmann, Phys. Rev. A {\bf 51}, 2443 (1995).

\bibitem{landau}
L. Landau and E. Lifshitz, {\it Statistical Physics},
(Pergamon, New York, 1958).

\bibitem{piezo}
T. Ikeda, {\it Fundamental of Piezoelectricity},
(Oxford University Press, 1990).

\bibitem{defS}
Given a frequency dependent operator ${\cal O}(\omega)$,
its spectrum ${\cal S}_{\cal O}(\omega)$ is defined through the relation
$\langle {\cal O}(\omega) {\cal O}(\omega') \rangle=
2\pi\delta(\omega+\omega')\,{\cal S}_{\cal O}(\omega)$.

\bibitem{diosi}
L. Diosi, Europhys. Lett. {\bf 22}, 1 (1993);
K. Jacobs, I. Tittonen, H. M. Wiseman and S. Schiller,
to appear in Phys. Rev. A, quant-ph/9902040.

\bibitem{duffin}
W. J. Duffin, {\it Electricity and Magnetism},
(McGraw-Hill, London, 1965).

\bibitem{defC}
Given two frequency dependent operators ${\cal O}_1(\omega)$, and ${\cal
O}_2(\omega)$,
their correlation function ${\cal C}_{{\cal O}_1{\cal O}_2}(\omega)$
is defined trough the relation
$\langle {\cal O}_1(\omega) {\cal O}_2(\omega') \rangle=
2\pi\delta(\omega+\omega')\,{\cal C}_{{\cal O}_1{\cal O}_2}(\omega)$.

\bibitem{tau95}
M. S. Taubman, H. M. Wiseman, D. E. McClelland, and
H. A. Bachor, J. Opt. Soc. Am. B {\bf 12}, 1792 (1995).

\bibitem{fblan}
H. M. Wiseman, Phys. Rev. A
{\bf 49}, 2133 (1994);
{\bf 49}, 5159 (1994);
{\bf 50}, 4428 (1994).

\bibitem{gala}
P. Galatola, L. A. Lugiato, M. G. Porreca, P. Tombesi
and G. Leuchs, Opt. Comm. {\bf 85}, 95 (1991).

\bibitem{pin99}
M. Pinard, Y. Hadjar and A. Heidmann,
Eur. Phys. J. D {\bf 7}, 107 (1999).

\bibitem{grav}
See e.g.,
{\it Quantum Optics, Experimental Gravitation and Measurement theory},
edited by P. Meystre and M. O. Scully, (Plenum, New York, 1983);
{\it Gravitational Wave Experiments}, edited by E. Coccia {\it et al.},
(World Scientific, Singapore, 1995).


\end{references}
\end{document}